\documentclass[useAMS,usenatbib,usegraphicx]{mn2e}

\usepackage{amsmath,amssymb,bm,upgreek,color}
\usepackage{mathrsfs}
\usepackage[latin1]{inputenc}
\usepackage{JournalNames}
\usepackage{upgreek}
\usepackage{enumerate}
\usepackage{wrapfig}
\usepackage{multicol}
\usepackage{lscape}
\usepackage{pifont}
\usepackage{subfigure}

\begin{document}

\title[Radiative MRI-turbulent Protoplanetary Discs]
{Vertical Structure and Turbulent Saturation Level in Fully Radiative
Protoplanetary Disc Models}

\author[M. Flaig et al.]{M. Flaig,$^1$ W. Kley$^1$ and R. Kissmann$^{2}$ \\
$^1$ Institut f\"ur Astronomie \& Astrophysik, Universit\"at T\"ubingen, 
Auf der Morgenstelle 10, 72076 T\"ubingen, Germany \\
$^2$ Institut f\"ur Astro- und Teilchenphysik, Universit\"at Innsbruck,
Technikerstra\ss{}e 25/8, 6020 Innsbruck, Austria}

\date{Accepted ??. Received ??; in original form ??}

\pagerange{\pageref{firstpage}--\pageref{lastpage}} \pubyear{????}

\maketitle

\label{firstpage}

\begin{abstract}
We investigate a massive ($\varSigma \sim 10\,000 \mbox{ g} \, \mbox{cm}^{-2}$
at 1 AU) protoplanetary disc model by means of 3D radiation magnetohydrodynamics
simulations.  The vertical structure of the disc is determined self-consistently
by a balance between turbulent heating caused by the MRI and radiative cooling.
Concerning the vertical structure, two different regions can be distinguished: A
gas-pressure dominated, optically thick midplane region where most of the
dissipation takes place, and a magnetically dominated, optically thin corona
which is dominated by strong shocks.  At the location of the photosphere, the
turbulence is supersonic ($M \sim 2$), which is consistent with previous results
obtained from the fitting of spectra of YSOs.  It is known that the turbulent
saturation level in simulations of MRI-induced turbulence does depend on
numerical factors such as the numerical resolution and the box size.  However,
by performing a suite of runs at different resolutions (using up to 64x128x512
grid cells) and with varying box sizes (with up to 16 pressure scaleheights in
the vertical direction), we find that both the saturation levels and the heating
rates show a clear trend to converge once a sufficient resolution in the
vertical direction has been achieved. 
\end{abstract}


\section{Introduction \& Motivation} 

Despite many years of observational and theoretical research, the physics of
protoplanetary accretion discs is still not well understood.  Significant
uncertainties remain both with respect to their physical composition as well as
their temporal evolution.  Their rather short lifetime of $\sim 10^6 \mbox{ to }
10^7 \, \mathrm{yr}$ \citep{Hartmann1998,SASpitzer2006} requires a physical
process that transfers angular momentum outwards efficiently so that the matter
can be accreted fast enough onto the star.  Usually it is assumed that some sort
of turbulence is acting in the disc, providing an effective viscosity that
drives accretion \citep[][hereafter SS]{ShakuraSyunyaev1973AnA}.  As of today,
the most promising mechanism for this scenario appears to be the {\it
magneto-rotational instability} (MRI), a linear instability that exists in
rotating, weakly magnetised shear flows and operates under very general
conditions \citep{BHRev1998,BalbusRev2003,Balbus2009}.

Apart from protoplanetary discs, the MRI plays an important role also in a
variety of other accreting systems, for example accretion discs in active
galactic nuclei, galactic accretion discs and core-collapse supernovae.
Although these systems cover very different physical regimes, they can be
modeled using similar numerical methods.  The most common approach is to employ
the shearing-box approximation, where one models a small patch of the accreting
disc in the co-moving frame using special ``shear-periodic'' boundary
conditions which are consistent with the background shear
flow~\citep[][hereafer HGB]{HGB1995}.  The shearing-box model has the advantage
that it is easy to set up, and that it is not very expensive computationally.
In contrast to the many works that make use of the shearing-box approximation,
there are only very few global simulations of protoplanetary discs
\citep{FromangNelson09,Flock10,Dzyurkevich10}, all employing an isothermal
equation of state.

The most interesting quantity that numerical simulations can provide is the
magnitude of the turbulent viscosity, which is usually measured in terms of the
alpha parameter~(SS, HGB).  Numerical simulations show that the MRI leads to
sustained turbulence which transports angular momentum outwards at a rate that,
for the case of protoplanetary discs (but maybe not for other systems), is
compatible with observations~\citep{KingEtAl07}.  The saturation level of the
MRI depends on numerical parameters such as the box
size~\citep{JohansenZonalFlow2009}, the numerical resolution~\citep[][herafter
SKH]{FromangPapaloizou07,SimonHawley2009,Davis2009,Shi2010} and the numerical
scheme~\citep{Balsara10}.  There is also a dependence on the physical parameters
viscosity and resisistivity~\citep{FromangEtAl2007} and, to a lesser
extent, also on the thermodynamics~(\nocite{FlaigEtAl2009}Flaig~et.~al.~2009,
herafter FK2; see also \nocite{TurnerEtAl03}Turner~et.~al.~2003).  When doing
MRI simulations, it is therefore very important to check the dependence of the
results on numerical factors and also to include as much of the relevant physics
as possible.

Protoplanetary discs are dense systems and therefore for a large part optically
thick.  This means that one should include radiation transport in order to
properly model the transport of radiation energy from the disc midplane, which
is constantly heated by turbulent dissipation, to the upper layers, where the
radiation escapes into space.  Including radiation transport is very important
in order to get the correct vertical structure, which will be determined by a
dynamic balance between turbulent heating and radiative cooling.  It is also
an important step towards the goal of comparing numerical simulations to actual
observations by modelling accretion disc spectra, and thereby deriving
constraints on physical parameters.

While such an attempt to include radiation transport in MRI-turbulent
protoplanetary disc simulations has not been reported so far, there exist
already several works which include radiative transfer in simulations of
accretion discs that are situated in other physical regimes~\citep[for
example][for the case of radiation dominated discs]{Turner2004,Krolik2007}.
The gas-pressure dominated simulation of ~\cite{HKS2006}, hereafter HKS, is most
closely related to our work, and will be the main source of reference.  In such
simulations it is possible to track in quite some detail the path that the
energy takes inside accretion discs (HKS) and it is possible to derive
observable consequences~\citep{BlaesEtAl2006}.

When dealing with protoplanetary discs one has to consider that these are not
only dense, but also cold, which means that for a large part they will only be
poorly ionised, leading to the formation of a ``dead zone''~\citep{Gammie1996}.
The effects of a finite resistivity have already been studied in isothermal
models~\citep[for example][]{SanoEtAl2000,TurnerDeadZone2008,Dzyurkevich10}.  In
the present work, we avoid this additional complication by placing our
simulation domain near to the star and choosing a surface density that is high
enough so that the turbulent heating provides a sufficient amount of
collisional ionisation, thereby justifying the assumption of ideal MHD.  In a
later work, we will extend our model by including a physical resisitvity,
thereby arriving at a protoplanetary disc model that includes all the relevant
physics in a self-consistent manner. 

The organisation of our paper is as follows: In \S 2 we describe our numerical
methods and the model setup.  In \S 3 the results of our simulations are
presented.  In \S 4 we conclude and provide an outlook on the further research
that is planned.

\section{Model description}

We apply the stratified shearing-box approximation, i.e. our simulations take
place in a co-moving rectangular box that is small enough so that local
Cartesian coordinates $(x,y,z)$ can be introduced, which correspond to the
radial, azimuthal and vertical coordinates, respectively.  The radial boundary
conditions are shear-periodic and the azimuthal boundary conditions are strictly
periodic~(HGB).  In the vertical direction we apply open boundary conditions
that allow outflow, but no inflow.  The simulation box is placed at a distance
of $1\,{\rm AU}$ from the central star which has one solar mass.  We choose a
surface density $\varSigma_0$ of order $\sim 10\,000\,\,{\rm g\, cm^{-2}}$,
which lies in between the value of $\varSigma_0 = 1700\,\,{\rm g\,cm^{-2}}$
given by the minimum mass solar nebula of~\cite{Hayashi1981} and the value of
$\varSigma_0 = 50500\,\,{\rm g\,cm^{-2}}$ according to the more recently
proposed solar nebula model by~\cite{Desch2007}.  With this choice of surface
density, the temperatures that come out of our model are of the order of
$1000-2000\,\mathrm{K}$ in the body of the disc, which allows us to describe the
the disc gas using the ideal MHD equations.  The gas is assumed to have solar
chemical composition, therefore we choose values of $\gamma=1.4$ for the
adiabatic index and $\mu=2.35$ for the mean molecular weight.

For protoplanetary discs, the stellar irradiation plays an important role in
determining the degree of ionisation in the disc, because stellar X-rays are one
of the main ionisation sources.  However, since for our choice of parameters the
disc can be considered to be already sufficiently ionised for the MRI to work,
we may safely neglect this factor.  At visible wavelengths, the stellar
irradiation will only be important in the very upper layers of a protoplanetary
disc, where according to passively heated models, a temperature inversion will
occur~\citep{ChiangGoldreich1997}.  In the present work we are mainly interested
in the temperature profile as it results from a balance between active heating
by turbulent dissipation and cooling by radiation transport.  Therefore we
exclude this region from our simulation domain and neglect the stellar
irradiation in the visible part of the spectrum also.  For reference, the basic
properties of our model are summarised in Table~\ref{table-model}.

\begin{table} 
\begin{minipage}{10cm}
	\begin{tabular}{lcc} 
\hline 
\hline 
	Parameter           & Symbol     & Value                        \\
\hline 
	Radial box size             & $L_x$      & ${\rm 0.08\, AU}$            \\
	Azimuthal box size          & $L_y$      & ${\rm 0.48\, AU}$            \\
	Vertical box size       & $L_z$ & ${\rm 0.8\, AU}$ - ${\rm 1.28\, AU}$ \\
	Mass of central star        & $M_\ast$   & $1\,M_\odot$                 \\
	Distance to central star    & $R_0$      & ${\rm 1 \, AU}$              \\
	Surface density        & $\varSigma_0$ & ${\rm 11\,356\,\, g\,cm^{-2}}$ \\
	Adiabatic index             & $\gamma$   & $1.4$                        \\
	Mean molecular weight       &$\mu$       & $2.35$                       \\
\hline 
	\end{tabular}
\end{minipage}
	\caption{\label{table-model}Summary of basic physical parameters of our
		model.  The vertical box size of $L_z=0.8\,{\rm AU}$ corresponds to a
		model which covers 10 pressure scale heights in the vertical direction,
		while $L_z=1.28\,{\rm AU}$ corresponds to a model with 16 scale
		heights.}
\end{table}

\subsection{Equations solved}

We calculate the dynamics of the disc gas by solving the equations of radiation
magnetohydrodynamics in conservative form.  We make the one-temperature
approximation (i.e. we assume thermal equilibrium between matter and radiation),
which means that we need not to solve an additional equations for the radiation
energy.
Radiative
diffusion is treated within the flux-limited diffusion approach.  Including the
source terms that arise in shearing box framework, the resulting set of
equations looks as follows:
\begin{subequations} \label{MHD-equations} 
	\begin{gather}
		\frac{\partial \rho}{\partial t} + \bm \nabla \cdot (\rho \bm v) = 0, \\
		\frac{\partial (\rho \bm v)}{\partial t} +
			\bm \nabla \cdot \left( \rho \bm v \bm v 
				+ P_\mathrm{tot} \mathsf{I}
				- \frac{\bm B \bm B}{4 \uppi} \right)
			= \bm f, \\
		\frac{\partial E_\mathrm{tot}}{\partial t} + \bm \nabla \cdot \left[
			(E_\mathrm{tot} + P_\mathrm{tot} ) \bm v
				- \frac{\bm B ( \bm B \cdot \bm v )}{4 \uppi} \right] 
			= \bm f \cdot \bm v - \bm \nabla \cdot \bm F, \\
		\frac{\partial \bm B}{\partial t} - \bm \nabla \times (\bm v \times \bm
B) = 0;
	\end{gather}
\end{subequations}
where the total pressure $P_\mathrm{tot} = p + B^2/8 \uppi$ is the sum of gas
pressure and magnetic pressure (neglecting the contribution arising from
radiation pressure), $\mathsf{I}$ is the identity matrix, $\bm f$ denotes the
sum of the gravitational forces and the inertial forces arising in the
shearing-box system, $E_\mathrm{tot} = p/(\gamma - 1) + \rho v^2/2 + B^2/8 \uppi
+ E_\mathrm{rad}$ is the total energy, $\bm F = -(\lambda c / \kappa \rho) \bm \nabla E_\mathrm{rad}$ is the
radiative energy flux, and the other symbols have their usual meaning.  Using
linearised expressions for the gravitational and inertial forces, $\bm f$
becomes (HGB):
\begin{equation} 
	\bm f = 3 \rho \varOmega^2 x \, \hat{\bm x}
		- \rho \varOmega^2 z \, \hat{\bm z} + 2 \rho \varOmega \bm v \times
		  \hat{\bm z},
\end{equation}
with $\varOmega$ the angular orbital frequency.  Note that by solving the total
energy equation, rather than the thermal energy equation, all dissipative losses
are automatically captured and transformed into gas internal energy.  In this
way the heating of the gas is consistent with the dissipation caused by the
turbulence.

In the optically thick protoplanetary discs, matter and radiation are closely
coupled and we can assume that they are at the same temperature, which means
that the radiation energy density is given by $E_\mathrm{rad}=aT^4$, which
closes our system of equations.  For the flux limiter $\lambda$ we apply the
formula given in Eq.~(28) of~\cite{LevermorePomraning1981}.  
\begin{figure} 
	\begin{minipage}{\textwidth}
		\includegraphics[width=0.45\textwidth]{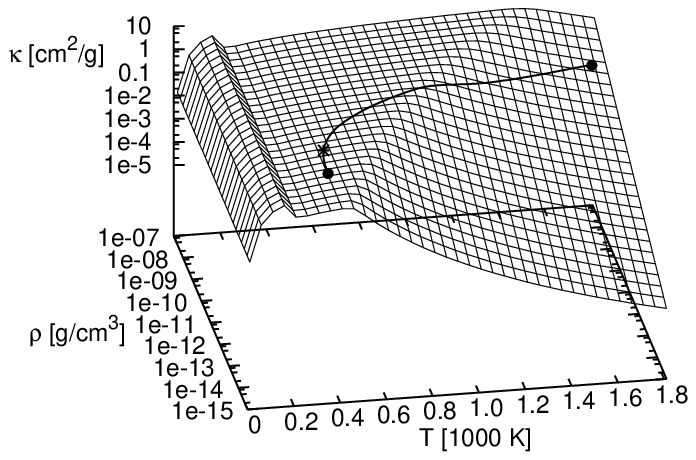}
	\end{minipage}
	\begin{picture}(0,0)(0,0)
		\put(59,139){\ding{192}}
		\put(70,140){\ding{193}}
		\put(100,145){\ding{194}}
		\put(180,152){\ding{195}}
		\put(101,85){{\sf S}}
		\put(199,122){{\sf M}}
	\end{picture}
	\caption{\label{Fig-Opacity}Opacity as a function of temperature and
density.  The thick curve gives the opacities for the typical densities and
temperatures found in our model \texttt{R7}, where \textsf{M} and \textsf{S}
correspond to the location of the midplane and the surface, respectively, while
the star $\star$ denotes the mean location of the photosphere.  For an
explanation of the different opacity regions \ding{192} to \ding{195} see the
text.}
\end{figure}
We use dust opacities $\kappa = \kappa(\rho,T)$ as described in
\cite{BellLin1994}, which are calculated as a function of the gas density and
temperature.  Fig.~\ref{Fig-Opacity} shows a plot of the opacity function for
the typical density and temperature range that we have in our simulations.
Four different regions can be distinguished: At the lowest temperatures, in
region \ding{192}, the dominant contribution to the opacity is due to ice
grains.  Since these are more efficient scatterers and absorbers at shorter
wavelengths, the opacity increases with temperature.  In region \ding{193},
ranging from $\sim 170$ to $\sim 200\,{\rm K}$, the ice grains melt, so the
opacity decreases.  Region \ding{194} is dominated by dust grains and the
opacity shows only a weak temperature dependence of the form $\kappa \propto
T^{0.5}$.  Finally, in region \ding{195}, the temperatures and densities are
sufficiently high for the dust grains to start melting, leading again to a
decrease of the opacity.

\subsection{Numerical code}

We use the \textsc{Cronos} magnetohydrodynamics code \citep{Kissmann2008}.  The
code has already been successfully applied to simulations of the turbulent
interstellar medium \citep{Kissmann2008}, coronal mass ejections
\citep{Pomoell2008} and simulations of MRI turbulence in accretion discs (FK2).
The code solves the ideal MHD equation using the conservative scheme described
in \cite{KurganovEtAl2001}.  Since we do not include any explicit dissipation,
the turbulent cascade is cut off at the grid scale, where kinetic and magnetic
energy are lost due to numerical errors and transformed into gas internal energy
by virtue of the conservative properties of the underlying numerical scheme.
The fact that the heating is mainly due to numerical effects means that there
will inevitably be some dependence on numerical parameters such as the
resolution and the details of the numerical scheme involved.  However, since
numerical errors are largest at locations where there are sharp gradients in
velocity and magnetic field, the numerical dissipation should at least
qualitatively resemble the actual physical dissipation~\citep[HKS,
][]{Krolik2007}.  For high enough resolutions, we expect the results to become
independent of numerical parameters.  It is therefore very important to perform
runs at different resolutions in order to check that the results are actually
converged.

The radiation transport is treated using operator splitting.  During the
radiation transport step, the equation $\partial (e_\mathrm{th} +
E_\mathrm{rad}) / \partial t = -\bm \nabla \cdot \bm F$ is solved, which, using
the relations $e_\mathrm{th} = \rho k_\mathrm{B} T / (\gamma - 1) \mu m_\mathrm{H}$ and
$E_\mathrm{rad}=aT^4$, can be transformed into the following diffusion equation
for the temperature:
\begin{equation} 
	(
		c_\mathrm{V} \rho
		+ 4 a T^3
	)
		\frac{\partial T}{\partial t} 
=
	\bm \nabla \cdot \frac{\lambda c}{\kappa \rho} 4 a T^3 \bm \nabla T
	,
	\label{radstep} 
\end{equation} 
where $c_\mathrm{V} = k_\mathrm{B} / (\gamma - 1) \mu m_\mathrm{H}$. In
order to avoid prohibitively short timesteps due to the presence of optically
thin regions, Eq.~\eqref{radstep} is solved implicitly (for more details on the
 implementation of the radiation transport, see the appendix). 

\subsection{\label{IniCond}Boundary \& initial conditions}

\begin{table*}
	\begin{tabular}{ccccccccc} 
\hline 
\hline 
		Name   & Resolution  & Box size & grid cells/$H_0$       & Orbits
      & TD     & $\langle \langle \alpha_\mathrm{Reyn} \rangle \rangle$ & $\langle \langle \alpha_\mathrm{Maxw} \rangle \rangle$ & $\langle \langle \alpha \rangle \rangle$ \\
\hline 
		\texttt{I6}  & 32x64x256 & $H_0 \times 6 H_0 \times 12 H_0$ & 21 & 200
      & isothermal   & $0.001\pm0.001$ & $0.003\pm0.004$ & $0.004\pm0.005$ \\                           
		\texttt{I5}  & 32x64x256 & $H_0 \times 6 H_0 \times 10 H_0$ & 25 & 80  
      & isothermal   & $0.003\pm0.002$ & $0.011\pm0.006$ & $0.014\pm0.008$ \\                          
		\texttt{I7}  & 32x64x448 & $H_0 \times 6 H_0 \times 14 H_0$ & 32 & 70  
      & isothermal   & $0.004\pm0.001$ & $0.013\pm0.003$ & $0.017\pm0.004$ \\                          
		\texttt{I6D} & 64x128x512& $H_0 \times 6 H_0 \times 12 H_0$ & 42 & 60  
      & isothermal   & $0.003\pm0.001$ & $0.012\pm0.005$ & $0.015\pm0.006$ \\
\hline 
		\texttt{R6}  & 32x64x256 & $H_0 \times 6 H_0 \times 12 H_0$ & 21 & 200
      & radiative    & $0.001\pm0.001$ & $0.004\pm0.003$ & $0.005\pm0.004$ \\   
		\texttt{R7} & 32x64x448 & $H_0 \times 6 H_0 \times 14 H_0$ & 32 & 150
      & radiative    & $0.003\pm0.002$ & $0.015\pm0.010$ & $0.018\pm0.013$ \\
		\texttt{R8}  & 32x64x512 & $H_0 \times 6 H_0 \times 16 H_0$ & 32 & 90  
      & radiative    & $0.004\pm0.002$ & $0.018\pm0.009$ & $0.022\pm0.011$ \\ 
		\texttt{R6D} & 64x128x512& $H_0 \times 6 H_0 \times 12 H_0$ & 42 & 60  
      & radiative    & $0.004\pm0.002$ & $0.017\pm0.008$ & $0.021\pm0.010$ \\                          
\hline 
		\end{tabular}
	\caption{\label{table-runs}Overview of the simulation runs that were
performed.  The box size (third column) is measured in scaleheights $H_0$
refering to the {\it initial} gas pressure at the midplane.  Since the pressure
at the midplane does not change very much, this corresponds very well to the
actual pressure scaleheight during the simulation.  The fourth column gives the
number of grid cells per scale height in the vertical direction.  The
fifth column tells the number of orbits for which the simulation has been run.
``TD'' (sixth column) denotes the the type of thermodynamics.  The last three
columns contain the time-averaged values of the turbulent stresses as well as
the standard deviation.}
\end{table*}

In the radial direction, shear-periodic boundary condtions are applied, which
are consistent with the shear flow in our local box (HGB).  The azimuthal
boundary conditions are periodic.  In the vertical direction, we use outflow
boundary conditions which extrapolate the values of the density and the velocity
in the outermost cell into the ghost zones.  In order to prevent inflow of mass,
reflective boundary conditions are applied instead whenever the velocity vector
points into the computational domain.  For the magnetic field, we apply vertical
field boundary conditions, where $B_x$ and $B_y$ are set to zero in the ghost
cells and $B_z$ is extrapolated from the outermost computational cell.  With
this set-up we do not need any (artificial or physical) resistivity near the
boundaries in order to prevent inconvenient effects.  At the boundaries, the
temperature is set to a small fixed value of $T_{\mathrm b} = {\rm 10\, K}$,
which is kept constant throughout the simulation.  This prescription for the
temperature allows for free emission of the radiation from the disc surface, and
it is well justified by the fact that in our models the gas at the boundary is
always optically thin.

As the initial condition 
we choose a state of
approximate vertical hydrostatic equilibrium:
\begin{equation} 
	\begin{array}{rcl}
		\rho(t=0)  &= &\rho_{\mathrm m} \exp(-z^2/2 H_0^2), \\
		\bm v(t=0) &= &\bm v_\mathrm{kep} + \delta \bm v, \\
		T(t=0)     &= &T_0, \\
		B(t=0)     &= &\bm \nabla \times \delta \bm A,
	\end{array}
\end{equation}
with the scale height $H_0 = c_{\mathrm{g0}} / \varOmega$, where
$c_{\mathrm{g0}} = \sqrt{k T_\mathrm{0} / \mu m_\mathrm H}$ is the initial gas
sound speed, and $\bm v_{\mathrm{kep}}  = -\frac32 \varOmega x
\, \hat{\bm y}$ is the background Keplerian shear flow.  For the initial
midplane density $\rho_\mathrm{m}$ and the initial temperature $T_\mathrm{0}$ we
choose the values $\rho_\mathrm{m} = {\rm 3.93 \cdot 10^{-9} \, g \, cm^{-3}}$
(corresponding to a midplane number density $n_\mathrm{m} = 10^{15} \, {\rm
cm^{-3}}$) and temperature $T_\mathrm{0} = {\rm 1500 \, K}$, respectively.  The surface
density is then $\varSigma_0 = {\rm 11356 \, g \, cm^{-2}}$.  In order to avoid
excessive shrinking or swelling of the disc during the inital phase, the initial
temperature $T_0$ has been chosen (by trial and error) such that it is roughly
the midplane temperature that comes out of the simulations for this particular
combination of $\varSigma_0$ and $R_0$.  Small random perturbations $\delta \bm v$
of order $10^{-3} \, c_\mathrm{g0}$ are added to the velocity.  Finally, the
initial magnetic field is derived from the following vector potential:
\begin{gather} \label{ZF-Field} 
	\delta \bm A = B_0 \frac{\sin(2 \uppi x/L_x)}{2 \uppi/L_x} \, \exp(-z^2/2 H_0^2) \,
	\hat{\bm y}.
\end{gather}
Note that the initial net magnetic flux in the computational domain is thus
zero, and this property is retained throughout the simulation to good accuracy.
The value of $B_0$ is taken such that it corresponds to a plasma beta of $\beta
= 100$ at the midplane.  With this particular choice of magnetic field, the
whole disc quickly becomes turbulent in less than 10 orbits.


\begin{figure*}
	\subfigure[\label{Butterfly}Horizontally averaged azimuthal magnetic field
component $B_y$ normalised to the initial magnetic field strength $B_0$.  The blue curve denotes the location of
the photosphere and the green curve is the location at which magnetic pressure
starts to exceed gas pressure.]{\includegraphics[width=0.8\textwidth,height=0.2\textwidth]{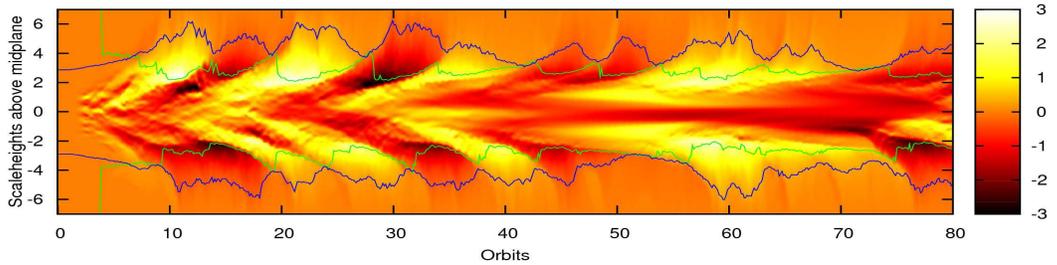}}
	\subfigure[\label{Energ-Hist}Volume averaged magnetic, turbulent kinetic and
thermal energy normalised to the initial gas pressure at the midplane.  The gas
internal energy is about 100 times the magnetic energy (note the different
scaling).]{\includegraphics[width=0.8\textwidth,height=0.2\textwidth]{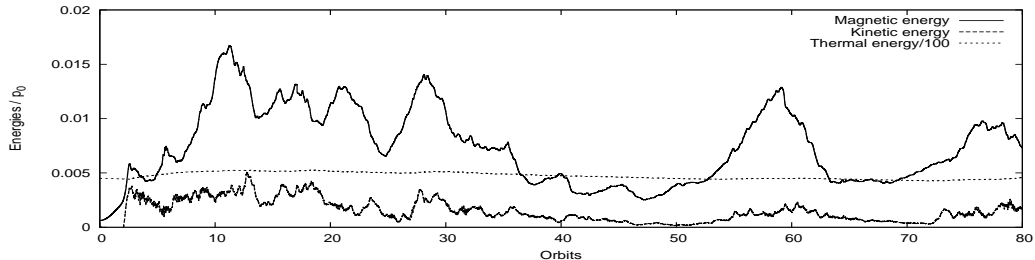}}
	\subfigure[\label{Alpha-Hist}Volume averaged turbulent stresses normalised to the mean gas pressure.]{\includegraphics[width=0.8\textwidth,height=0.2\textwidth]{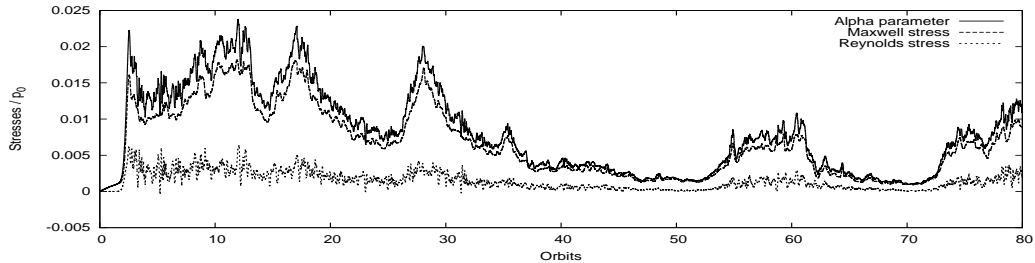}}
	\subfigure[\label{Teff-Hist}Photospheric temperature at
top and bottom boundary.]{\includegraphics[width=0.8\textwidth,height=0.2\textwidth]{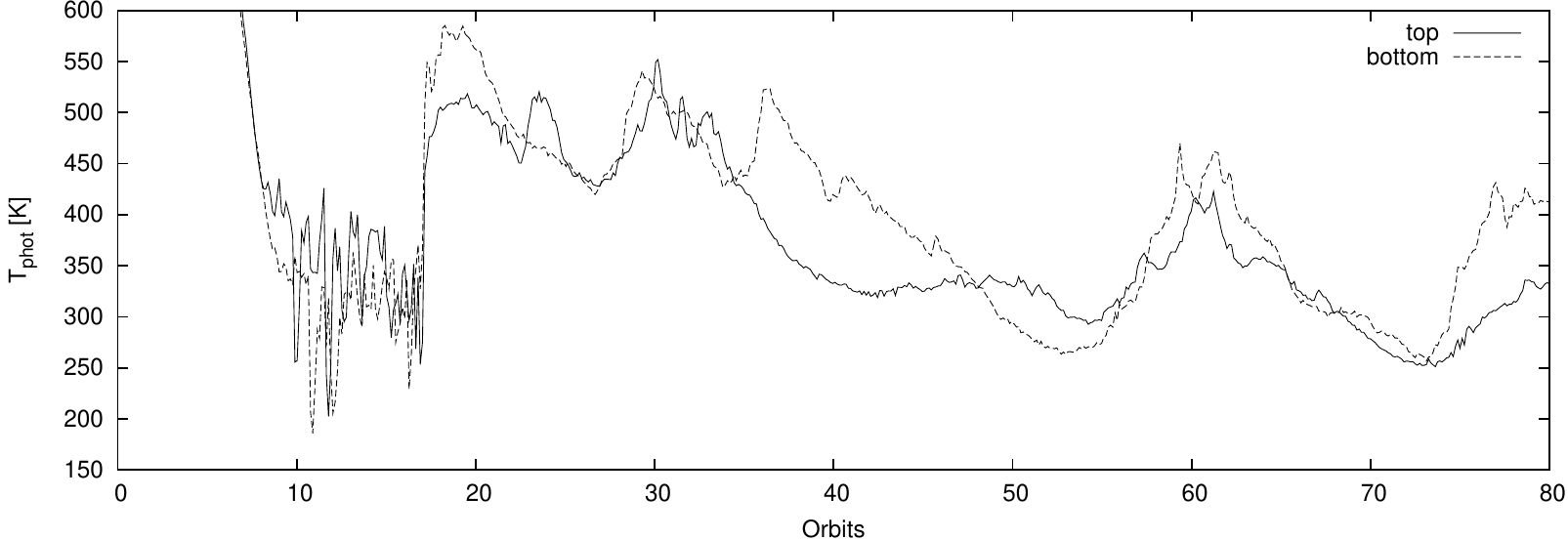}}
	\caption{\label{Time-Hist}Time history of various physical variables for
model \texttt{R7} from 0 to 80 orbits illustrating the changes in turbulent
activity and the correlations between stresses, magnetic field strength and
luminosity of the disc.  For further explanation see the text.}
\end{figure*}

\subsection{Definition of mean quantities}

Since we want to investigate the vertical structure of the disc, we are
especially interested in horizontally averaged quantities.  Therefore, for any
quantity $q$, we define the radially and azimuthally averaged value $\langle q
\rangle_{xy}$ as
\begin{equation} \label{z-gemittelt}
	\langle q \rangle_{xy}
	= \frac{1}{L_x L_y} \int \!\!\! \int q \, \mathrm dx \, \mathrm dy.
\end{equation}
In addition to this, we will also use the volume average $\langle q
\rangle$, which is defined as
\begin{equation} \label{volumenmittel}
	\langle q \rangle = \frac{1}{L_x L_y L_z} \int \!\!\! \int \!\!\! \int q \,
		\mathrm dx \, \mathrm dy \, \mathrm dz.
\end{equation}
The definitions~\eqref{z-gemittelt} and ~\eqref{volumenmittel} are of course to
be understood in a discretised sense.  

The most important factor that determines the disc's temporal evolution is the
amount of angular momentum transport that takes place inside the disc.  It is
related to the $r-\phi$ component of the total stress~\citep{BHRev1998}, which,
in the local shearing-box frame, reads
\begin{equation}
    T_{xy} = \rho v_x \delta v_y - \frac{B_x B_y}{4 \uppi},
\end{equation}
where $\delta v_y = v_y + \frac32 \varOmega x$.  $T_{xy}$ can be decomposed as
the sum of Reynolds stress $T_{\rm Reyn} = \rho v_x \delta v_y$ and Maxwell
stress $T_{\rm Maxw} = -B_x B_y / 4 \uppi$.

Following the $\alpha$ prescription of SS, we normalise the stresses to the mean
gas pressure.  We thus define the volume-averaged alpha parameter as 
\begin{equation} \label{def-alpha}
	\langle \alpha \rangle 
	= 
	\langle \alpha_\mathrm{Reyn} \rangle + \langle
	\alpha_\mathrm{Maxw} \rangle,
\end{equation}
where
\begin{equation}
	\langle \alpha_\mathrm{Reyn} \rangle = \langle T_{\rm Reyn} \rangle / \langle P \rangle
	\mbox{ and }
	\langle \alpha_\mathrm{Maxw} \rangle = \langle T_{\rm Maxw} \rangle / \langle P \rangle
\end{equation}
are the volume-averaged Reynolds and Maxwell stresses normalised by the mean gas
pressure.

\section{\label{Sec-Simulations}Simulations} 

Table~\ref{table-runs} provides an overview of the simulations that were
performed.  We perform simulations with different box sizes and resolutions.
The lowest resolution chosen is $32\times64\times256$ (Model \texttt{R6}).  We
also perform a simulation at double the resolution in all directions (simulation
\texttt{R6D}) and two simulations (\texttt{R7} and \texttt{R8}) with a larger
box size in the vertical direction and a vertical resolution that is in between
model \texttt{R6} and \texttt{R6D}, since recent results indicate that it is the
resolution in the vertical direction that is critical in determining the value
of the turbulent saturation level (SKH).  In order to see how the results from
the radiative simulations compare to simulations which use an isothermal
equation of state, we also perform a set of simulations (\texttt{I5},
\texttt{I6}, \texttt{I7} and \texttt{I6D}) which have identical physical
parameters as the radiative simulations but which use an isothermal equation of
state where the temperature is constant kept constant at the initial value of
$T_0$.

\subsection{Time history}

Starting from the initial state described in Sec.~\ref{IniCond} the MRI starts
to grow quickly and the non-linear state is reached already after a few orbits.
After ten orbits the whole disc is fully turbulent and remains turbulent for the
whole course of the simulation.  Concerning the time history we now concentrate
on model \texttt{R7}.

The turbulent state is not time-steady, but there are periods of high turbulent
activity which are followed by periods where the disc is less active.  This
behaviour can be observed in Fig.~\ref{Time-Hist}, where various physical
quantities are plotted as function of time for the first 80 orbits.
Fig.~\ref{Butterfly} shows the horizontally averaged azimuthal component of the
magnetic field (which is the dominant component).  During an active phase the
magnetic field is lifted upwards, leading to the typical butterfly structure.
Recently it has been claimed that it is the undulatory
Parker instability that drives the updwelling of the magnetic field (SKH).  As
one can estimate from Fig.~\ref{Butterfly}, one cycle lasts between 10 and 20
orbits.  
Often the changes in turbulent activity are accompanied by magnetic field
reversals in one or both sides of the disc.  Unlike, for example, the solar
butterfly diagram, the field reversals shown in Fig.~\ref{Butterfly} do not
follow a regular pattern and there is no strict correlation between the magnetic
polarities in both sides of the disc.

In Fig.~\ref{Butterfly} we also plot the location of the photosphere (defined as
the location where the optical depth equals unity) and the position of the
magnetosphere (defined as the location where the magnetic pressure starts to
exceed the gas pressure).  Just between two active phases, the locations of the
magnetosphere and the photosphere coincide, so the magnetically dominated region
is optically thin.  During an active phase, the disc expands due to the
stronger magnetic forces and the photosphere is pushed outwards.  At the same
time, the magnetosphere is pushed inwards due to the higher magnetisation caused
by stronger turbulence.  Therefore, for much of the time, the magnetically
dominated region is for a large part optically thick.  The same phenomenon has
also been reported in the gas-pressure dominated simulation of HKS, although the
physical regime is quite different.
 
We plot the time history of the energies (thermal energy, turbulent kinetic
energy and magnetic energy) in Fig.~\ref{Energ-Hist}.  During active periods the
magnetic energy is much larger then during a quiet period.  The turbulent
kinetic energy (that is the total kinetic energy minus the kinetic energy of the
background shear flow) is also larger during active phases.  In contrast to
this, the long-term changes in the thermal energy are much smaller than the
up-and-down variations, which shows that we have indeed reached thermal
equilibrium.

The turbulent stresses that determine the angular momentum transport in the disc
are shown in Fig.~\ref{Alpha-Hist}.  The angular momentum transport is dominated
by the Maxwell stress which is about four to six times larger than the Reynolds
stress.  This ratio is consistent with previous results for stratified
boxes~\citep[for example][]{Stone1996}.  From a comparison between
Fig.~\ref{Energ-Hist} and Fig.~\ref{Alpha-Hist} it is evident that the magnetic
energy and the alpha parameter are strongly correlated.  Indeed, it is known
that in shearing-box calculations, the time-dependence of the alpha parameter
can be fitted by a formula of the form $\langle \alpha \rangle \propto \langle
B^2 \rangle + {\rm const.}$ \citep[see][]{Brandenburg2008PS}.

\begin{figure} 
	\includegraphics[width=0.45\textwidth]{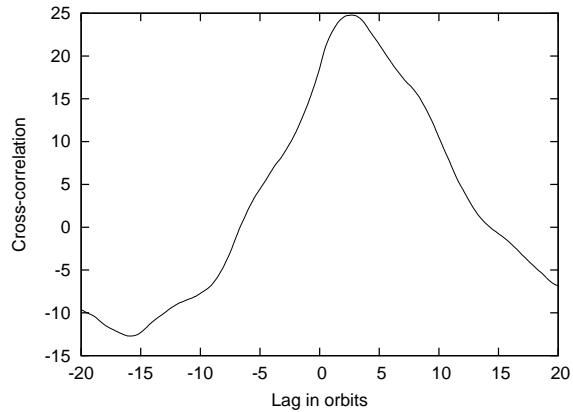}
	\caption{\label{Fig-crosscorr}Cross-correlation between alpha
stress and photospheric temperature for model \texttt{R7}.}
\end{figure}

\begin{figure} 
	\includegraphics[width=0.45\textwidth]{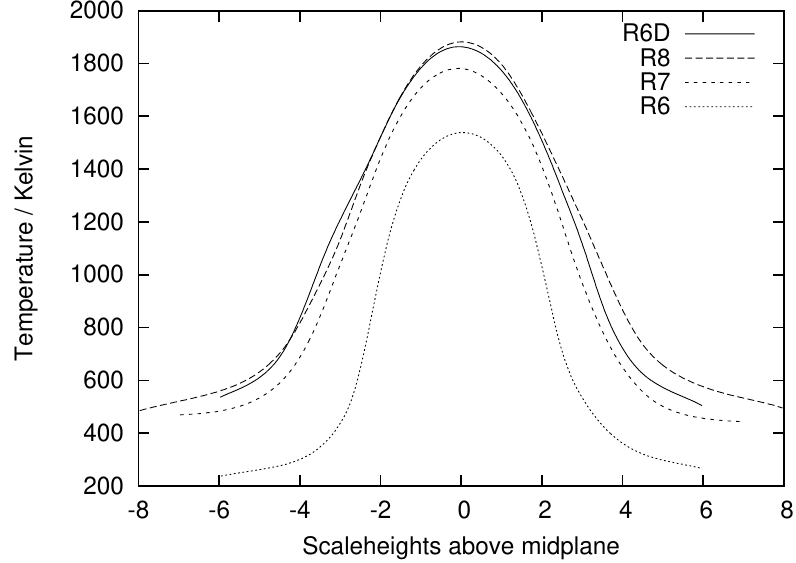}
	\caption{\label{Mean_Temp}Temperature profiles obtained for the
radiative models.  Averages have been taken from 20 orbits until the end of the
simulation.}
\end{figure}

Finally, we may ask how the observational appearance of the disc will change due
to phases of varying turbulent activity.  To address this question, we plot in
Fig.~\ref{Teff-Hist} the photospheric temperature (i.e. the temperature at the
location where the optical depth equals unity).  If we compare
Fig.~\ref{Teff-Hist} with Fig.~\ref{Alpha-Hist}, it is obvious that in general a
high degree of turbulent activity leads to a higher flux of radiation through
the disc's boundaries.  The increased flux then balances the increased turbulent
heating, so that the thermal energy content in the disc remains nearly constant.
This observation can be made more quantitative by looking at the
cross-correlation between alpha-parameter and photospheric temperature
(Fig.~\ref{Fig-crosscorr}):
\begin{equation}
	C(\tau) 
=
	\int [\langle \alpha \rangle (t-\tau) - \langle \langle \alpha \rangle
		\rangle ] 
	[T_\mathrm{phot}(t) - \langle T_\mathrm{phot} \rangle] \, \mathrm dt.
\end{equation}
Here, $\langle \alpha \rangle$ denotes the volume-averaged alpha parameter as
defined in Eq.~\eqref{def-alpha} and $T_\mathrm{phot}$ is the arithmetic mean of
the photospheric temperatures at top and bottom.  The lag $\tau$ between
photospheric temperature and alpha parameter (as inferred from the peak in the
correlation function) is about 2-3 orbits.  We can compare this to the
radiative diffusion timescale $\tau_\mathrm{rad} = H_0^2/D_\mathrm{rad}$, where
the radiative diffusion coefficient $D_\mathrm{rad}$ is given by
\begin{equation} 
	D_\mathrm{rad} 
	=
	4 a c T^3 / 3 c_\mathrm{V}
	\kappa \rho^2
\end{equation} 
(cf.  the appendix).  When taking typical values $\rho = 2.0 \cdot 10^{-9} \,
\mathrm{g\,cm^{-3}}$ and $T=1700\,\mathrm{K}$ for the density and the
temperature in the region near the midplane (see Fig.~\ref{Mean_Temp} and Fig.~\ref{Mean_Rho}
later in the paper), we get $\tau_\mathrm{rad} \sim 2.5$ orbits, so the lag
between stress and photospheric temperature agrees well with the radiative
diffusion timescale.  Concerning the turbulent transport coefficient
$D_\mathrm{turb} = v_z^2 / \varOmega$, we find that near the midplane
$D_\mathrm{turb}$ is typically about one order of magnitude smaller than
$D_\mathrm{rad}$, so the energy transport by turbulent gas motions is 
negligible compared to the energy transport by radiative diffusion.


\begin{figure*} 
	\subfigure[Magnetic field strength $|\bm B|$ measured in Gauss.]{%
		\begin{minipage}{0.3\textwidth}
			\begin{center}
				\includegraphics[width=\textwidth,height=2.0\textwidth]{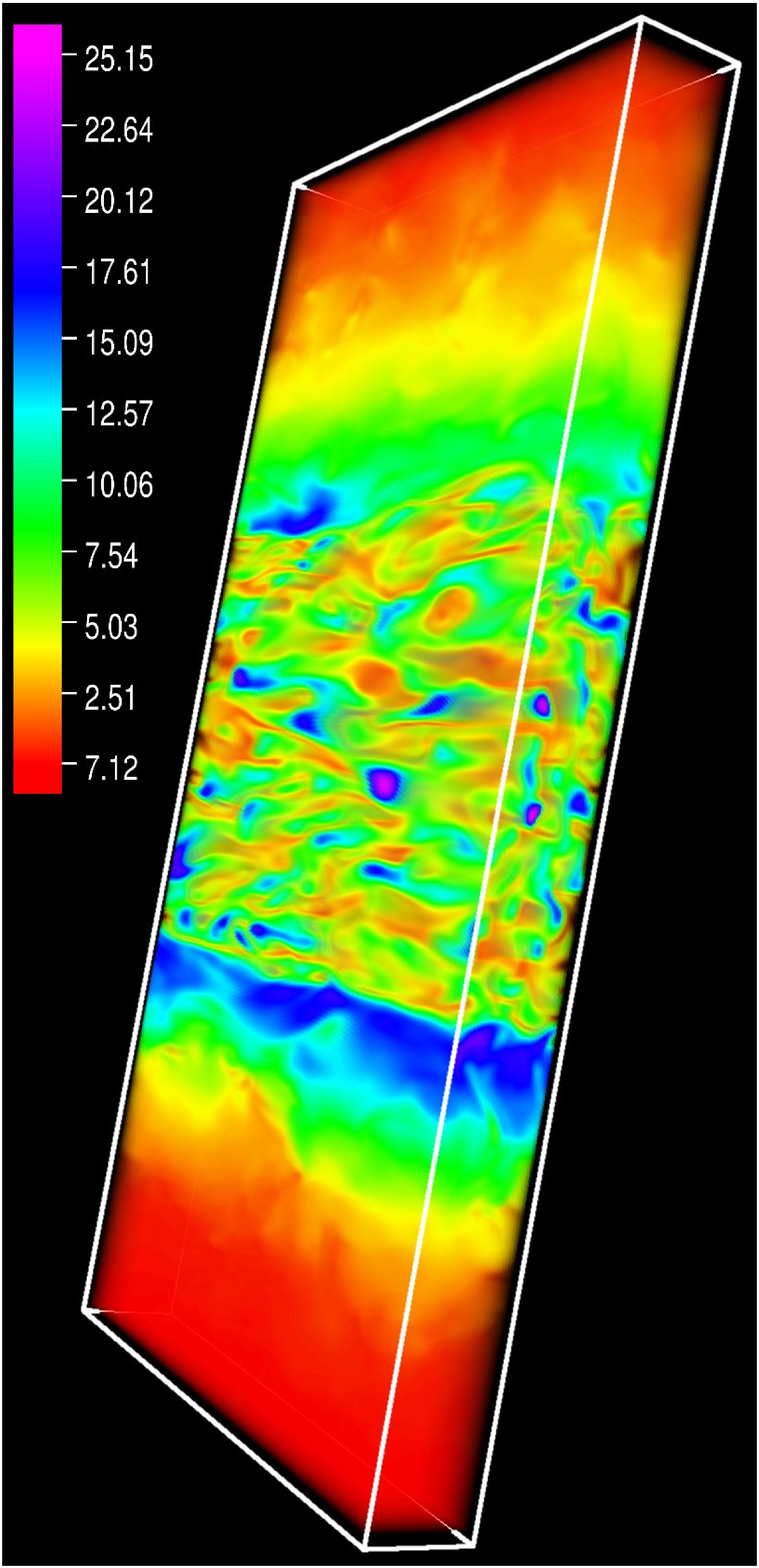}
			\end{center}
		\end{minipage}
	}
	\begin{picture}(0,0)(0,0)
		\put(-145,-10){\textcolor{white}{\textsf{Gauss}}}
		\put(-145,-25){\textcolor{white}{$|\bm B|$}}
	\end{picture}
	\subfigure[Vorticity $\omega=|\bm \nabla \times \bm v|$ in units of angular 
		orbital frequency $\varOmega$.]{%
		\begin{minipage}{0.3\textwidth}
			\begin{center}
				\includegraphics[width=\textwidth,height=2.0\textwidth]{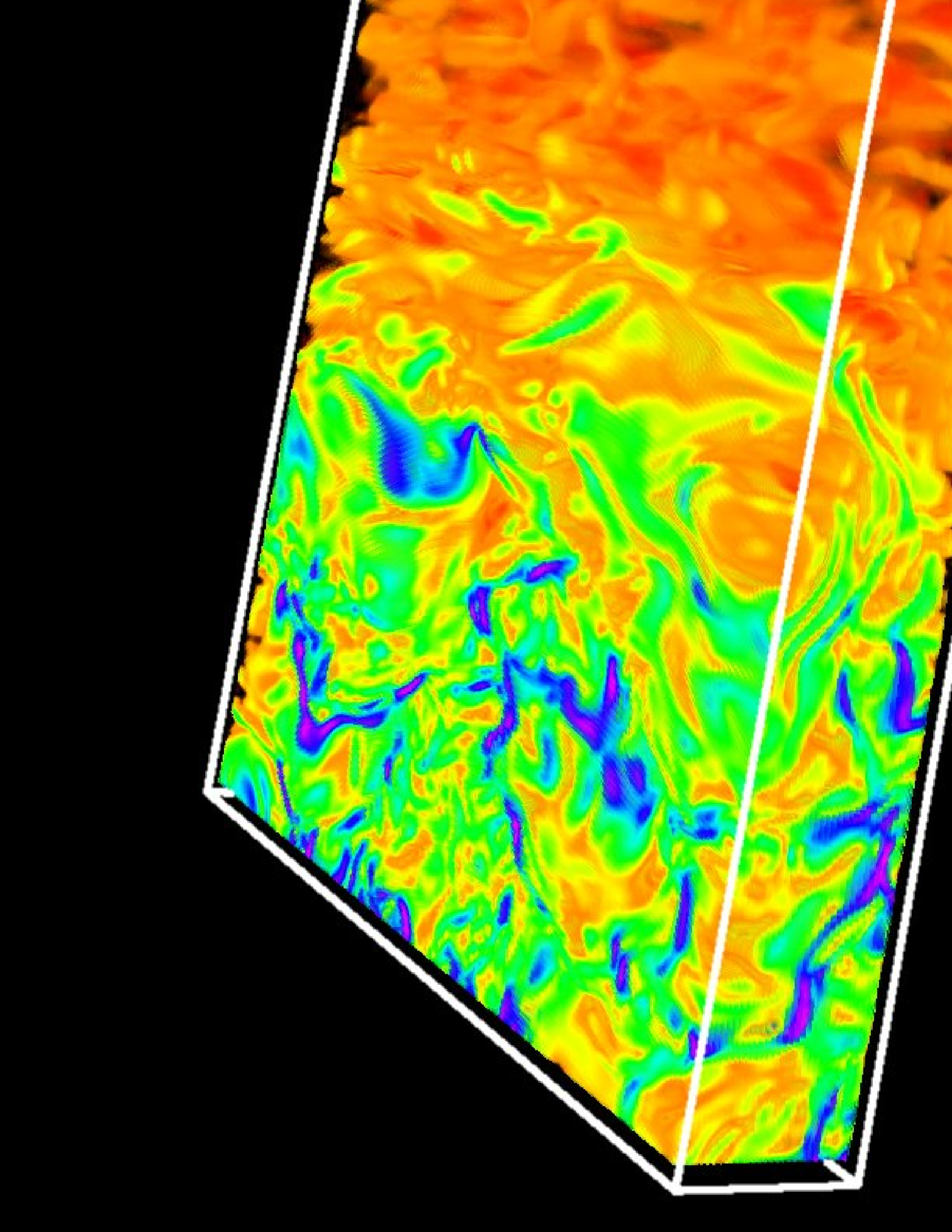}
			\end{center}
		\end{minipage}
	}
	\begin{picture}(0,0)(0,0)
		\put(-148,-20){\textcolor{white}{$\frac{\displaystyle \omega}%
			{\displaystyle \varOmega}$}}
	\end{picture}
	\subfigure[Ratio of photon mean free path to the length over which
the temperature varies.]{
		\begin{minipage}{0.3\textwidth}
			\begin{center}
				\includegraphics[width=\textwidth,height=2.0\textwidth]{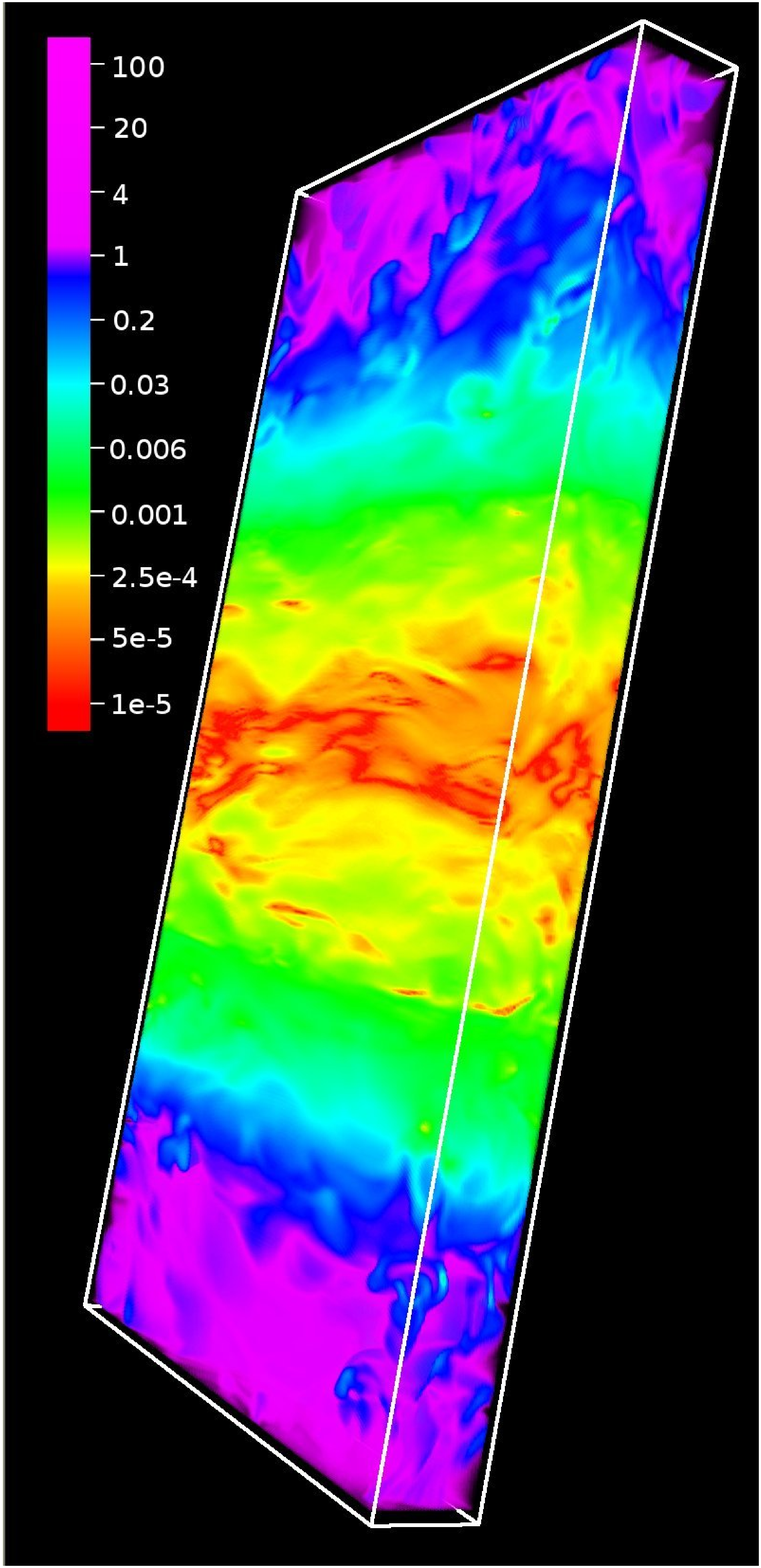}
			\end{center}
		\end{minipage}
	}
	\begin{picture}(0,0)(0,0)
		\put(-155,-10){\textcolor{white}{$\frac{\displaystyle 
		4 |\partial_z T|}{\displaystyle \kappa \rho T}$}}
	\end{picture}
	\begin{picture}(0,0)(100,100)
		\put(-278,-45){\textcolor{white}{\scalebox{1.2}{$x$}}}
		\put(-255,80){\textcolor{white}{\scalebox{1.2}{$z$}}}
		\put(-350,-30){\textcolor{white}{\scalebox{1.2}{$y$}}}
		\put(-119,-45){\textcolor{white}{\scalebox{1.2}{$x$}}}
		\put(-96,80){\textcolor{white}{\scalebox{1.2}{$z$}}}
		\put(-183,-30){\textcolor{white}{\scalebox{1.2}{$y$}}}
		\put(42,-45){\textcolor{white}{\scalebox{1.2}{$x$}}}
		\put(65,80){\textcolor{white}{\scalebox{1.2}{$z$}}}
		\put(-20,-30){\textcolor{white}{\scalebox{1.2}{$y$}}}
	\end{picture}
	\caption{\label{Fig-Snap}Snapshots from model \texttt{R6D} illustrating the
two-layer structure of the disc.  The plots in panels (a) and (b) have been
taken at $t=20$ orbits, the plot in panel (c) has been taken at $t=21$ orbits.
For further comments see the text. (Graphics have been produced using the
\textsc{Vapor} visualisation package.)}
\end{figure*}

\subsection{Vertical structure}

To get an impression of how the turbulent state looks like, the reader may first
have a look at Fig.~\ref{Fig-Snap}, where 3D snapshots of some physical
quantities are shown for the high-resolution model \texttt{R6D}.  Panel (a)
depicts the magnetic field strength measured in Gauss.  In panel (b) a plot of 
the vorticity measured in units of the angular orbital frequency $\varOmega$ is
shown.  Panel (c) contains a plot of the quantity 
\begin{equation}
	R_z
	= \frac{|\partial_z E_\mathrm{rad} |}{\kappa \rho E_\mathrm{rad}}
	= \frac{4 |\partial_z T |}{\kappa \rho T}
\end{equation}
which is the ratio of the photon mean free path $\ell_\mathrm{phot} = 1/\kappa
\rho$ to the typical length over which the radiation energy (or the
temperature) varies.

Concerning the vertical structure, two regions
with very different physical properties can be disinguished:  On the one hand,
there is the region near the midplane (the region inside the first
three scaleheights).  Here, the magnetic field is approximately uniform and of
the order of several Gauss.  The turbulence is subsonic (see Sec.~\ref{turb-vel}
below) and many intertwined vortices can be observed.  Since $R_z \ll 1$ the
photons are diffusing in this region.  On the other hand, there is the corona
(the region outwards from 4 scaleheights) which exhibits markedly
different features: Here, the magnetic field drops sharply, the flow is
supersonic and characterised by strong shocks, and the photons are to a large
part better described as free streaming rather than diffusing. 

Due to the properties of the shearing-box setup, on average the physical state
will appear homogeneous with respect to the radial and the azimuthal direction.
This suggests to consider horizontal averages of physical quantities.  However,
due to the turbulent nature of the flow, the horizontal average of some variable
at one instant in time shows large fluctuations and sometimes strong asymmetries
between the two halves of the disc.  It is therefore necessary to perform
averages over many orbits in order to achieve a smooth and symmetric
vertical profile.

\subsubsection{Density \& temperature profile}

One of the most interesting quantities that our radiative models are able to
provide is the self-consistently calculated temperature profile.  We plot the
mean temperature profiles for our radiative models in Fig.~\ref{Mean_Temp}.
Except for the low resolution model \texttt{R6} all the radiative models yield
nicely consistent temperature profiles.  In the body of the disc, the
temperature profile resembles an inverted parabola and becomes flat in the
optically thin regions in the upper layers of the disc.  Near the disc midplane,
\, the temperature is about $1800$ K.  The mean position of the photosphere is
located at about 5 scaleheights away from the midplane.  The corresponding
photospheric temperatures are of the order of $500-600$~K.  According to the
well-known formula
\begin{equation}
	\dot M
	= \frac{8 \uppi}{3} \frac{\sigma T^4 R_0^3}{G M_\odot}
\end{equation}
this would correspond to an accretion rate $\dot M$ of order $10^{-5} \, M_\odot
\, \mbox{yr}^{-1}$.  The accretion rate can also be estimated from the alpha
parameter according to the formula given in \cite{Pringle1981ARAnA}
\begin{equation}
	\dot M
	= 3 \uppi \alpha c_\mathrm{g} H \varSigma_0.
\end{equation}
Using $c_\mathrm{g} \approx c_{\mathrm g0}$ as well as $H \approx H_\mathrm{0}$,
and plugging in a value of $\alpha=0.02$ for the alpha parameter (see
Sec.~\ref{TurbSat}), we get again
an accretion rate of order $10^{-5} \, M_\odot \, \mbox{yr}^{-1}$, so the two
results are nicely consistent.



\begin{figure} 
	\includegraphics[width=0.45\textwidth]{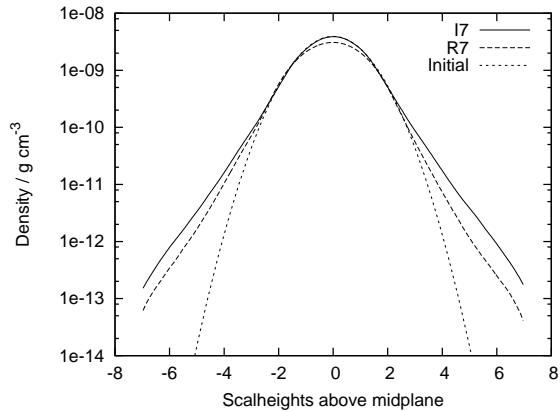}
	\caption{\label{Mean_Rho}Density profile for models \texttt{I7} and
\texttt{R7}, averaged from 20 orbits until the end of the simulation.  For
comparison, the initial gaussian density profile has also been plotted
(short-dashed line).}
\end{figure}

In Fig.~\ref{Mean_Rho} we compare the density profile obtained with
model \texttt{R7} to the profile of the isothermal model \texttt{I7} and also
to the initial profile.  Away from the midplane the density profiles flatten due
to the additional magnetic support.  Since for the radiative model the
temperature decreases outwards, the density in the upper layers is lower as
compared to the isothermal model.


\subsubsection{Vertical support \& Stresses}

\begin{figure} 
\includegraphics[width=0.45\textwidth]{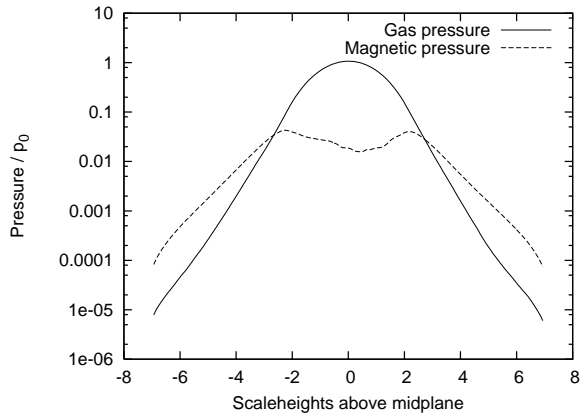}
\caption{\label{Fig-PressureSupport}Gas pressure and magnetic pressure for
model \texttt{R7}.  Averages have been taken from 20 orbits until the end of
the simulation.}
\end{figure}

In our model, the radiation pressure is small compared to the gas pressure
almost everywhere.  This means that concerning the vertical support of the disc
against gravity, radiation pressure plays no role, leaving only gas pressure and
magnetic forces.  As can be inferred from Fig.~\ref{Fig-PressureSupport}, where
gas and magnetic pressure are plotted as a function of height, the midplane
region inside the first three scaleheights is gas-pressure dominated.  The
magnetic pressure is approximately constant in the midplane region with a slight
increase outwards.  Outside the midplane region, the magnetic pressure declines
exponentially, but not as steep as the gas pressure.  As a consequence, the
disc's corona is magnetically dominated. 


\begin{figure} 
	\includegraphics[width=0.45\textwidth]{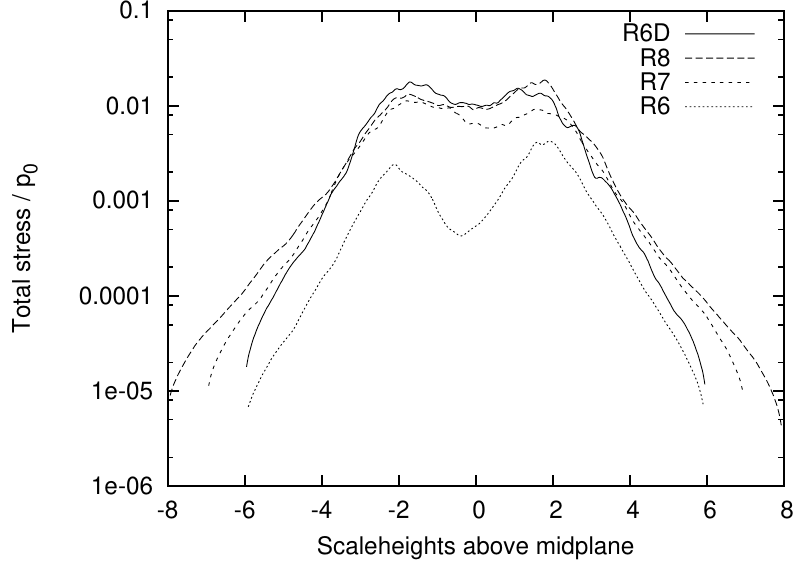}
	\caption{\label{Fig-Mean_Stress}Turbulent stresses as a function of height
for the radiative models.  Averages have been taken from 20 orbits until the end of the simulation.}
\end{figure}

The turbulence is not only important with regard to the transport of angular
momentum, but also insofar as it constantly creates new magnetic field and heats
the disc gas; balancing the losses of thermal energy and magnetic field 
and thus preventing the collapse of the disc.  We can
determine the strength of the turbulence at different locations by measuring the
turbulent stresses.  This is done in Fig.~\ref{Fig-Mean_Stress}, where vertical
profiles of the alpha parameter are plotted for the radiative models.  Note that
except for the low resolution model \texttt{R6} all profiles are very similar,
indicating convergence.  We encounter a similar picture as in
Fig.~\ref{Fig-PressureSupport} for the case of the magnetic pressure: Inside the
gas-pressure dominated midplane region the stress profiles are roughly constant
with a slight increase outwards (with the exception of the low-resolution model
\texttt{R6}, where the stresses drop noticably at the midplane).  Outside of the
midpane region the stress profiles decline exponentially.  This means that
almost all the angular momentum transport happens in the midplane region and is
there almost independent of height.  In the corona, angular momentum transport
is negligible.

\subsubsection{\label{turb-vel}Turbulent velocities}

\begin{figure} 
	\includegraphics[width=0.45\textwidth]{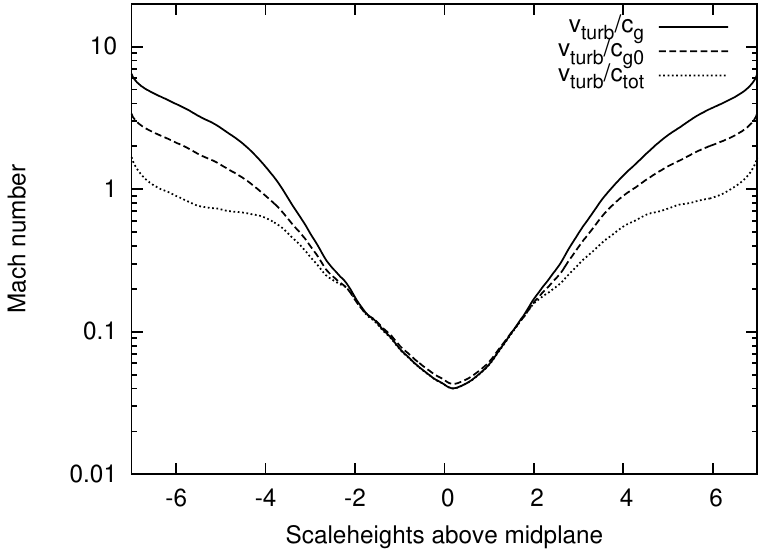}
	\caption{\label{Fig-vturb}Turbulent velocities for model \texttt{R7}, where
$c_{\mathrm g}$ denotes the gas sound speed, $c_{\mathrm g0}$ is the initial gas
sound speed at the midplane and $c_\mathrm{tot}$ is the total sound speed
(including gas and magnetic pressure).   Averages have been taken from 20 to 60
orbits.}
\end{figure}

We now look at the velocity distribution in the disc. In Fig.~\ref{Fig-vturb} we
plot the horizontally averaged turbulent velocity $\bm v_\mathrm{turb} = \bm v -
\bm v_\mathrm{kep}$ normalised by the gas sound speed as a function of height.
In the midplane region, the turbulence is subsonic,  while in the corona it
becomes highly supersonic, exceeding Mach 5 at the boundaries.  For comparison
with other works we also plot the velocity normalised to the initial isothermal
sound speed at the midplane and the velocity normalised to the total sound speed
$c_\mathrm{tot}$, where $c_\mathrm{tot}^2 = (p + B^2/8 \uppi) /\rho$.  Our
results are in agreement with the isothermal simulations done by
\cite{MillerStone00} who report Mach numbers of about two and with HKS, who
report Mach numbers (with respect to the total sound speed) between one and two.

From Fig.~\ref{Fig-Snap}, middle panel, where a snapshot of the vorticity is plotted, on can
appreciate the highly dynamic and tangled nature of the velocity field.  In the
subsonic midplane region many intertwined vortices can be distinguished.
Vortices have been proposed as a means of trapping dust particles, thereby
helping the formation of larger bodies by enhancing the number of collisions and
slowing down the spiraling into the star~\citep{BargeSom1995,Tanga1996}.
However, the vortices in our simulations are rather short-lived, usually lasting
much shorter than one orbit, so particle trapping in the MRI-generated vortices
in the midplane region will not be efficient.

In observations, the turbulence will show itself in the form of a broadening of
spectral lines due to the velocity dispersion induced by it.  At the mean
location of the photosphere the turbulent velocities are about Mach 2-3 which
implies a significant effect on the line widths.  The detection of CO overtone
emission in young stellar objects (YSOs) provides a useful diagnostic tool to
detect turbulent line broadening.  The reason for this is that the near overlap
of CO transitions near the $v=2-0$ band allows a separation of the local
broadening (e.g. turbulence) from the macrobroadening (caused for example by a
disc wind).  In the case of several YSOs, there is indeed substantial empirical evidence
for supersonic turbulent line broadening of the magnitude that we find in our
simulations \citep{Najita1996,Carr2004,Huegel}.

\subsection{\label{TurbSat}Turbulent saturation level}

As has already been remarked in the introduction, in numerical simulations of
MRI tubulence, the turbulent saturation level is influenced by a number of
numerical factors.  By performing a suite of simulations with different box
sizes an at different resolutions, we are able to gauge the strength of the
influence of these numerical parameters.

\begin{figure} 
\includegraphics[width=0.45\textwidth]{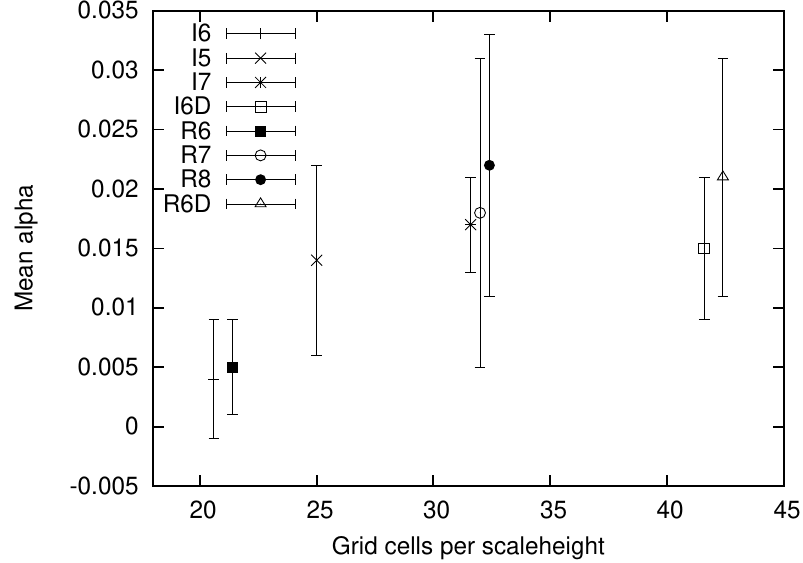}
\caption{\label{Fig-ResStudy}Time-averaged value of the alpha parameter as a
function of the number of grid cells in the {\it vertical} direction per scale
height for all models.  The ``error bars'' correspond to the deviation from the
mean value.}
\end{figure}

In Fig.~\ref{Fig-ResStudy} we plot the values of the alpha parameter as given in
Table~\ref{table-model}.  When increasing the resolution from 32x64x256 (models
\texttt{I6} and \texttt{R6}), to the double resolution of 64x128x512 (models
\texttt{I6D} and \texttt{R6D}), the values of the stresses increase
considerably.  A similar increase of the stresses is observed also when only the
resolution in the vertical direction is increased (models \texttt{I5},
\texttt{I7}, \texttt{R7} and \texttt{R8}), which is analogous to what has been
found in the radiative simulations of SKH.  A look at Fig.~\ref{Fig-Mean_Stress}
suggests that it is especially the region near the midplane that is not
sufficiently resolved, since the stress drops there noticably, while in the
better resolved models it does not.

The results depicted in Fig.~\ref{Fig-ResStudy} suggest that once a resolution
of about 30 grid cells per scaleheight in the vertical direction has been
achieved, the results will no longer significantly depend on the numerical
resolution.  All models except \texttt{I6} and \texttt{R6} are consistent with
an $\alpha$-value of about $\alpha \sim 0.015-0.02$, with the isothermal models
yielding somewhat lower stresses that are closer to the lower end of $\alpha
\sim 0.015$, while the stresses found in the radiative models are grouped around
the value $\alpha \sim 0.02$.  No tendency is found for the turbulent saturation
level to change with respect to neither the vertical box size nor the resolution
in the radial and azimuthal directions.


As a consequence of the lower stress, the heating in the low-resolution model
\texttt{R6} is also smaller than compared to the other radiative models, leading
to significantly lower temperatures in this model.  Apart from this, the
temperature profile for model \texttt{R6} is qualitatively similar to the other
models.  

As has already remarked before, the heating does intrinsically depend on
numerical effects.  However, as can be seen from Fig.~\ref{Mean_Temp}, the
temperature profiles for the better resolved models (\texttt{R7}, \texttt{R8}
and \texttt{R6D}) are very similar, suggesting that the heating rates are also
converged.

\section{Conclusion \& Outlook} \label{Sec-Conclusion}

We have performed 3D radiative MHD simulations of MRI-turbulent protoplanetary
discs.  Including radiation transport allows us to obtain a self-consistent
picture of the vertical structure which results from a dynamic balance between
turbulent heating and radiative cooling.  Such simulations do thus contain a
greater level of realism as compared to previous isothermal simulations.
Furthermore, radiative simulations are an important step towards bringing
numerical simulations into contact with observations.

In our models two regions can be distinguished: A gas-pressure dominated
midplane region where the magnetic field and stresses are approximately constant
on average and where almost all the angular momentum transport and dissipation
of turbulent energy take place, and a magnetically dominated corona where the
stresses and the magnetic field are much smaller than in the midplane region.
The position of the photosphere is highly variable due to changing turbulent
activity.  For most of the time, a large part of the magnetically dominated part
of the disc is optically thick. Only at times when the turbulent activity is low
do the positions of the photosphere and the magnetosphere coincide.

Since the turbulent saturation level in simulations of MRI turbulence may depend
on the numerical resolution, and since in our simulations the heating of the
disc is to a large part due to numerical dissipation, it is especially important
to check what influence numerical parameters such as the resolution or the box
size have on the outcome of the simulation.  By running models at different
resolutions and with different box sizes, we are able to verify previous results
that it is the resolution in the vertical direction that is critical for
achieving convergence of the turbulent stresses.  We find that once a sufficient
resolution in the vertical direction has been achieved ($\sim 30$ grid cells per
scaleheight), neither the saturation level nor the heating rates do
significantly change when increasing the resolution further.  No trend is found
for either of these to quantities to change with the vertical box size.


For our model parameters, where we chose to simulate a massive protoplanetary
disc, the turbulent heating alone already proved to be sufficient to justify the
assumption of ideal MHD.  The logical next step will be to include resistivity
into the model and choose parameters that are appropriate to a ``standard''
protoplanetary disc model.  When including both non-ideal MHD and radiation
transport, it will be possible to do truly self-consistent simulations that do
not only include the proper thermodynamics, but also a self-consistently evolved
dead zone.

\subsection*{Acknowledgements}
This research has been supported in part by the Deutsche Forschungsgemeinschaft
DFG through grant DFG Forschergruppe 759 ``The Formation of Planets: The
Critical First Growth Phase''.  Computational resources were provided by the
H\"ochstleistungsrechenzentrum Stuttgart (HLRS) and the High Performance
Computing Cluster of the University of T\"ubingen.  We thank Neal Turner for
very helpful discussions.  We thank an anonymous referee for providing valuable
suggestions that helped improve the paper.

\appendix 
\section{Implementation of Radiation Transport}
\label{App-RT-Implementation}

In this appendix we describe the implementation of radiation transport into the
\textsc{Cronos} MHD code.  By adding the thermal energy equation
\begin{equation}
	\frac{\partial e_\mathrm{th}}{\partial t}
	+ \bm \nabla \cdot ( e_\mathrm{th} \bm v )
	=
	- p \bm \nabla \cdot \bm v
	- \kappa_\mathrm{P} \rho (4 \uppi B - c E_\mathrm{rad})
\end{equation}
and the radiation energy equation 
\begin{equation}
	\frac{\partial E_\mathrm{rad}}{\partial t}
	+ \bm \nabla \cdot ( E_\mathrm{rad} \bm v )
	=
	- \frac{E_\mathrm{rad}}{3} \bm \nabla \cdot \bm v
	+ \kappa_\mathrm{P} \rho (4 \uppi B - c E_\mathrm{rad})
	- \bm \nabla \cdot \bm F,
\end{equation}
we obtain an equation that describes the evolution of the combined
(gas+radiation) energy:
\begin{equation}
	\frac{\partial ( e_\mathrm{th} + E_\mathrm{rad} )}{\partial t}
	+ \bm \nabla \cdot [( e_\mathrm{th} + E_\mathrm{rad} ) \bm v ]
	=
	- p_\mathrm{tot} \bm \nabla \cdot \bm v
	- \bm \nabla \cdot \bm F,
\end{equation}
where in general $p_\mathrm{tot} = p + E_\mathrm{rad}/3$.  However, since in our
simulations the radiation pressure is small compared to the gas pressure, we
neglect it and set $p_\mathrm{tot} = p$.  Splitting off the radiation diffusion
part and making use of the flux-limited diffusion approximation
\citep{LevermorePomraning1981}, the radiation transport step becomes:
\begin{equation}
	\frac{\partial (e_\mathrm{th} + E_\mathrm{rad})}{\partial t}
	=
	-\bm \nabla \cdot \frac{\lambda c}{\kappa \rho} \bm \nabla E_\mathrm{rad}.
\end{equation}
Using the ideal gas law $e_\mathrm{th} = c_\mathrm{V} \rho T$ and assuming
thermal equilibrium between matter and radiation, $E_\mathrm{rad} = a T^4$, we
close our system of equations and end up with the following diffusion equation for the radiation energy:
\begin{equation} \label{RadEnEq}
	\left(\frac{e_\mathrm{th}}{4E_\mathrm{rad}} + 1\right) \frac{\partial E_\mathrm{rad}}{\partial t}
	=
	-\bm \nabla \cdot \frac{\lambda c}{\kappa \rho} \bm \nabla E_\mathrm{rad}.
\end{equation}
From this we can estimate the timescale to reach thermal balance according to
$\tau \sim H^2/D_\mathrm{rad}$, where $H$ is the scale height and
\begin{equation}
	D_\mathrm{rad} = \left( \frac{e_\mathrm{th}}{4 E_\mathrm{rad}} + 1 \right)^{-1} 
	\frac{\lambda c}{\kappa \rho}.
\end{equation}
Note that in the radiation pressure dominated case, $D_\mathrm{rad}$ is simply
given by $D_\mathrm{rad} = \lambda c / \kappa \rho$.  However, for the gas
pressure dominated case that we are considering, we actually have
$D_\mathrm{rad} = 4 E_\mathrm{rad} \lambda c / e_\mathrm{th} \kappa \rho$, which
means that $\tau$ is bigger by a factor of $4 E_\mathrm{rad}/e_\mathrm{th}$ and
it takes thus much longer to reach thermal balance.  

When working in the gas-pressure dominated regime, it is natural to replace the
radiation energy $E_\mathrm{rad}$ in Eq.~\eqref{RadEnEq} by $aT^4$ and then
solve the following diffusion equation for the temperature:
\begin{equation} \label{TempEnEq}
	( c_\mathrm{V} \rho + 4aT^3) \frac{\partial T}{\partial t}
	=
	-\bm \nabla \cdot \frac{\lambda c}{\kappa \rho} 4 a T^3 \bm \nabla T.
\end{equation}
This equation is straightforwardly discretised in the following manner:
\begin{align*}
    &T^{n+1}_{ijk} 
    =
    T^n_{ijk} + \frac{\Delta t}{c_\mathrm{V} \rho_{ijk}^n + 4 a (T_{ijk}^n)^3} \Bigg\{ \\
    &\frac{1}{\Delta x^2}
    \left[
        D^n_{i+\frac12jk} \Delta T^{n+1}_{i+\frac12jk} -
		D^n_{i-\frac12jk} \Delta T^{n+1}_{i-\frac12jk}
    \right] + \\
    &\frac{1}{\Delta y^2}
    \left[
        D^n_{ij+\frac12k} \Delta T^{n+1}_{ij+\frac12k} -
		D^n_{ij-\frac12k} \Delta T^{n+1}_{ij-\frac12k}
    \right] + \\
    &\frac{1}{\Delta z^2}
    \left[
        D^n_{ijk+\frac12} \Delta T^{n+1}_{ijk+\frac12} -
		D^n_{ijk-\frac12} \Delta T^{n+1}_{ijk-\frac12}
    \right] \Bigg\}
    ,
\end{align*}
where $\Delta T^{n+1}_{i+\frac12jk} = T^{n+1}_{i+1jk} - T^{n+1}_{ijk}$,
\begin{equation}
	D^n_{i+\frac12jk} = \frac12 
		\left[ 
			\left(\frac{\lambda c}{\kappa \rho} 4 a T^3 \right)_{i+1jk}^n
			+ \left(\frac{\lambda c}{\kappa \rho} 4 a T^3 \right)_{ijk}^n
		\right],
\end{equation}
and so on.  The resulting matrix equation, which involves a diagonally dominant
matrix, can then be solved by any standard matrix solver.  For the calculations
reported in this paper we used the method of successive overrelaxation.

\label{SecAppend}
\bibliography{Paper}

\begin{thebibliography}{}

\bibitem[\protect\citeauthoryear{{Balbus}}{{Balbus}}{2003}]{BalbusRev2003}
{Balbus} S.~A.,  2003, \araa, 41, 555

\bibitem[\protect\citeauthoryear{{Balbus}}{{Balbus}}{2009}]{Balbus2009}
{Balbus} S.~A.,  2009, ArXiv e-prints

\bibitem[\protect\citeauthoryear{Balbus \& Hawley}{Balbus \&
  Hawley}{1998}]{BHRev1998}
Balbus S.~A.,  Hawley J.~F.,  1998, Rev. Mod. Phys., 70, 1

\bibitem[\protect\citeauthoryear{{Balsara} \& {Meyer}}{{Balsara} \&
  {Meyer}}{2010}]{Balsara10}
{Balsara} D.~S.,  {Meyer} C.,  2010, ArXiv e-prints

\bibitem[\protect\citeauthoryear{{Barge} \& {Sommeria}}{{Barge} \&
  {Sommeria}}{1995}]{BargeSom1995}
{Barge} P.,  {Sommeria} J.,  1995, \aap, 295, L1

\bibitem[\protect\citeauthoryear{{Bell} \& {Lin}}{{Bell} \&
  {Lin}}{1994}]{BellLin1994}
{Bell} K.~R.,  {Lin} D.~N.~C.,  1994, \apj, 427, 987

\bibitem[\protect\citeauthoryear{{Blaes}, {Davis}, {Hirose}, {Krolik} \&
  {Stone}}{{Blaes} et~al.}{2006}]{BlaesEtAl2006}
{Blaes} O.~M.,  {Davis} S.~W.,  {Hirose} S.,  {Krolik} J.~H.,    {Stone} J.~M.,
   2006, \apj, 645, 1402

\bibitem[\protect\citeauthoryear{Brandenburg}{Brandenburg}{2008}]{Brandenburg2%
008PS}
Brandenburg A.,  2008, Physica Scripta, p. 014016

\bibitem[\protect\citeauthoryear{{Carr}, {Tokunaga} \& {Najita}}{{Carr}
  et~al.}{2004}]{Carr2004}
{Carr} J.~S.,  {Tokunaga} A.~T.,    {Najita} J.,  2004, \apj, 603, 213

\bibitem[\protect\citeauthoryear{{Chiang} \& {Goldreich}}{{Chiang} \&
  {Goldreich}}{1997}]{ChiangGoldreich1997}
{Chiang} E.~I.,  {Goldreich} P.,  1997, \apj, 490, 368

\bibitem[\protect\citeauthoryear{{Davis}, {Stone} \& {Pessah}}{{Davis}
  et~al.}{2009}]{Davis2009}
{Davis} S.~W.,  {Stone} J.~M.,    {Pessah} M.~E.,  2009, ArXiv e-prints

\bibitem[\protect\citeauthoryear{{Desch}}{{Desch}}{2007}]{Desch2007}
{Desch} S.~J.,  2007, \apj, 671, 878

\bibitem[\protect\citeauthoryear{{Dzyurkevich}, {Flock}, {Turner}, {Klahr} \&
  {Henning}}{{Dzyurkevich} et~al.}{2010}]{Dzyurkevich10}
{Dzyurkevich} N.,  {Flock} M.,  {Turner} N.~J.,  {Klahr} H.,    {Henning} T.,
  2010, \aap, 515, A70+

\bibitem[\protect\citeauthoryear{{Flaig}, {Kissmann} \& {Kley}}{{Flaig}
  et~al.}{2009}]{FlaigEtAl2009}
{Flaig} M.,  {Kissmann} R.,    {Kley} W.,  2009, \mnras, 394, {1887 (FK2)}

\bibitem[\protect\citeauthoryear{{Flock}, {Dzyurkevich}, {Klahr} \&
  {Mignone}}{{Flock} et~al.}{2010}]{Flock10}
{Flock} M.,  {Dzyurkevich} N.,  {Klahr} H.,    {Mignone} A.,  2010, \aap, 516,
  A26+

\bibitem[\protect\citeauthoryear{{Fromang} \& {Nelson}}{{Fromang} \&
  {Nelson}}{2009}]{FromangNelson09}
{Fromang} S.,  {Nelson} R.~P.,  2009, \aap, 496, 597

\bibitem[\protect\citeauthoryear{{Fromang} \& {Papaloizou}}{{Fromang} \&
  {Papaloizou}}{2007}]{FromangPapaloizou07}
{Fromang} S.,  {Papaloizou} J.,  2007, \aap, 476, 1113

\bibitem[\protect\citeauthoryear{{Fromang}, {Papaloizou}, {Lesur} \&
  {Heinemann}}{{Fromang} et~al.}{2007}]{FromangEtAl2007}
{Fromang} S.,  {Papaloizou} J.,  {Lesur} G.,    {Heinemann} T.,  2007, \aap,
  476, 1123

\bibitem[\protect\citeauthoryear{{Gammie}}{{Gammie}}{1996}]{Gammie1996}
{Gammie} C.~F.,  1996, \apj, 457, 355

\bibitem[\protect\citeauthoryear{{Hartmann}, {Calvet}, {Gullbring} \&
  {D'Alessio}}{{Hartmann} et~al.}{1998}]{Hartmann1998}
{Hartmann} L.,  {Calvet} N.,  {Gullbring} E.,    {D'Alessio} P.,  1998, \apj,
  495, 385

\bibitem[\protect\citeauthoryear{{Hawley}, {Gammie} \& {Balbus}}{{Hawley}
  et~al.}{1995}]{HGB1995}
{Hawley} J.~F.,  {Gammie} C.~F.,    {Balbus} S.~A.,  1995, \apj, 440, {742
  (HGB)}

\bibitem[\protect\citeauthoryear{{Hayashi}}{{Hayashi}}{1981}]{Hayashi1981}
{Hayashi} C.,  1981, Progress of Theoretical Physics Supplement, 70, 35

\bibitem[\protect\citeauthoryear{{Hirose}, {Krolik} \& {Stone}}{{Hirose}
  et~al.}{2006}]{HKS2006}
{Hirose} S.,  {Krolik} J.~H.,    {Stone} J.~M.,  2006, \apj, 640, {901 (HKS)}

\bibitem[\protect\citeauthoryear{H\"ugelmeyer}{H\"ugelmeyer}{2009}]{Huegel}
H\"ugelmeyer S.,  2009, PhD thesis, Georg-August-Universit\"at G\"ottingen,
  G\"ottingen

\bibitem[\protect\citeauthoryear{{Johansen}, {Youdin} \& {Klahr}}{{Johansen}
  et~al.}{2009}]{JohansenZonalFlow2009}
{Johansen} A.,  {Youdin} A.,    {Klahr} H.,  2009, \apj, 697, 1269

\bibitem[\protect\citeauthoryear{{King}, {Pringle} \& {Livio}}{{King}
  et~al.}{2007}]{KingEtAl07}
{King} A.~R.,  {Pringle} J.~E.,    {Livio} M.,  2007, \mnras, 376, 1740

\bibitem[\protect\citeauthoryear{{Kissmann}, {Kleimann}, {Fichtner} \&
  {Grauer}}{{Kissmann} et~al.}{2008}]{Kissmann2008}
{Kissmann} R.,  {Kleimann} J.,  {Fichtner} H.,    {Grauer} R.,  2008, \mnras,
  391, 1577

\bibitem[\protect\citeauthoryear{{Krolik}, {Hirose} \& {Blaes}}{{Krolik}
  et~al.}{2007}]{Krolik2007}
{Krolik} J.~H.,  {Hirose} S.,    {Blaes} O.,  2007, \apj, 664, 1045

\bibitem[\protect\citeauthoryear{Kurganov, Noelle \& Petrova}{Kurganov
  et~al.}{2001}]{KurganovEtAl2001}
Kurganov A.,  Noelle S.,    Petrova G.,  2001, SIAM J. Sci. Comput., 23, 707

\bibitem[\protect\citeauthoryear{{Levermore} \& {Pomraning}}{{Levermore} \&
  {Pomraning}}{1981}]{LevermorePomraning1981}
{Levermore} C.~D.,  {Pomraning} G.~C.,  1981, \apj, 248, 321

\bibitem[\protect\citeauthoryear{{Miller} \& {Stone}}{{Miller} \&
  {Stone}}{2000}]{MillerStone00}
{Miller} K.~A.,  {Stone} J.~M.,  2000, \apj, 534, 398

\bibitem[\protect\citeauthoryear{{Najita}, {Carr}, {Glassgold}, {Shu} \&
  {Tokunaga}}{{Najita} et~al.}{1996}]{Najita1996}
{Najita} J.,  {Carr} J.~S.,  {Glassgold} A.~E.,  {Shu} F.~H.,    {Tokunaga}
  A.~T.,  1996, \apj, 462, 919

\bibitem[\protect\citeauthoryear{{Pomoell}, {Vainio} \& {Kissmann}}{{Pomoell}
  et~al.}{2008}]{Pomoell2008}
{Pomoell} J.,  {Vainio} R.,    {Kissmann} R.,  2008, \solphys, 253, 249

\bibitem[\protect\citeauthoryear{{Pringle}}{{Pringle}}{1981}]{Pringle1981ARAnA}
{Pringle} J.~E.,  1981, \araa, 19, 137

\bibitem[\protect\citeauthoryear{{Sano}, {Miyama}, {Umebayashi} \&
  {Nakano}}{{Sano} et~al.}{2000}]{SanoEtAl2000}
{Sano} T.,  {Miyama} S.~M.,  {Umebayashi} T.,    {Nakano} T.,  2000, \apj, 543,
  486

\bibitem[\protect\citeauthoryear{{Shakura} \& {Syunyaev}}{{Shakura} \&
  {Syunyaev}}{1973}]{ShakuraSyunyaev1973AnA}
{Shakura} N.~I.,  {Syunyaev} R.~A.,  1973, \aap, 24, {337 (SS)}

\bibitem[\protect\citeauthoryear{{Shi}, {Krolik} \& {Hirose}}{{Shi}
  et~al.}{2010}]{Shi2010}
{Shi} J.,  {Krolik} J.~H.,    {Hirose} S.,  2010, \apj, 708, 1716 (SKH)

\bibitem[\protect\citeauthoryear{{Sicilia-Aguilar}, {Hartmann}, {Calvet},
  {Megeath}, {Muzerolle}, {Allen}, {D'Alessio}, {Mer{\'{\i}}n}, {Stauffer},
  {Young} \& {Lada}}{{Sicilia-Aguilar} et~al.}{2006}]{SASpitzer2006}
{Sicilia-Aguilar} A.,  {Hartmann} L.,  {Calvet} N.,  {Megeath} S.~T.,
  {Muzerolle} J.,  {Allen} L.,  {D'Alessio} P.,  {Mer{\'{\i}}n} B.,  {Stauffer}
  J.,  {Young} E.,    {Lada} C.,  2006, \apj, 638, 897

\bibitem[\protect\citeauthoryear{{Simon}, {Hawley} \& {Beckwith}}{{Simon}
  et~al.}{2009}]{SimonHawley2009}
{Simon} J.~B.,  {Hawley} J.~F.,    {Beckwith} K.,  2009, \apj, 690, 974

\bibitem[\protect\citeauthoryear{{Stone}, {Hawley}, {Gammie} \&
  {Balbus}}{{Stone} et~al.}{1996}]{Stone1996}
{Stone} J.~M.,  {Hawley} J.~F.,  {Gammie} C.~F.,    {Balbus} S.~A.,  1996,
  \apj, 463, 656

\bibitem[\protect\citeauthoryear{{Tanga}, {Babiano}, {Dubrulle} \&
  {Provenzale}}{{Tanga} et~al.}{1996}]{Tanga1996}
{Tanga} P.,  {Babiano} A.,  {Dubrulle} B.,    {Provenzale} A.,  1996, Icarus,
  121, 158

\bibitem[\protect\citeauthoryear{{Turner}}{{Turner}}{2004}]{Turner2004}
{Turner} N.~J.,  2004, \apjl, 605, L45

\bibitem[\protect\citeauthoryear{{Turner} \& {Sano}}{{Turner} \&
  {Sano}}{2008}]{TurnerDeadZone2008}
{Turner} N.~J.,  {Sano} T.,  2008, \apjl, 679, L131

\bibitem[\protect\citeauthoryear{{Turner}, {Stone}, {Krolik} \&
  {Sano}}{{Turner} et~al.}{2003}]{TurnerEtAl03}
{Turner} N.~J.,  {Stone} J.~M.,  {Krolik} J.~H.,    {Sano} T.,  2003, \apj,
  593, 992

\end{thebibliography}
\label{lastpage}
\end{document}